\documentclass[apj,numberedappendix]{emulateapj}
\usepackage{natbib,apjfonts}

\begin{document}

\title{A Spreading Layer Origin for Dwarf Nova Oscillations}

\author{Anthony L. Piro\altaffilmark{1}
and Lars Bildsten\altaffilmark{1,2}}
\altaffiltext{1}{Department of Physics, Broida Hall, University of California,
Santa Barbara, CA 93106; piro@physics.ucsb.edu}

\altaffiltext{2}{Kavli Institute for Theoretical Physics,
Kohn Hall, University of California,
Santa Barbara, CA 93106; bildsten@kitp.ucsb.edu}

\begin{abstract}

  Dwarf nova outbursts often show coherent ($Q\sim10^4-10^6$)
sinusoidal oscillations with the largest pulsed fraction
in the extreme
ultraviolet. Called dwarf nova oscillations (DNOs),
they have periods of $P\approx3-40\textrm{ s}$ and scale with
luminosity as $P\propto L^{-\beta}$ with $\beta\approx0.1-0.2$.
We propose that DNOs may be produced by nonradial oscillations
in a thin hydrostatic layer of freshly accreted material, the
``spreading layer'' (SL), at the white dwarf (WD) equator. This
would naturally explain a number of key properties of DNOs,
including their frequency range, sinusoidal
nature, sensitivity to accretion rate, and why they
are only seen during outburst. In support of this hypothesis
we construct a simple model that treats the SL as a cavity containing
shallow surface waves, each with the same radial structure, but split
into three different modes denoted by their azimuthal wavenumber, $m$.
The $m=0$ latitudinally propagating
mode best matches the periods and scalings associated with
most DNOs, and DNOs with periods shorter than the WD Keplerian
period are explained by the $m=-1$ prograde mode. We
also predict a third set of oscillations, produced by the $m=1$
retrograde mode, and show its expected dependence on
accretion rate.
% Our new idea for explaining DNOs will hopefully
% motivate more  sophisticated models of nonradial oscillations in the SL
% so as to make a detailed test of our hypothesis.

\end{abstract}

\keywords{accretion, accretion disks ---
	novae, cataclysmic variables ---
	stars: oscillations ---
	white dwarfs}

\section{Introduction}

  Rapid oscillations are often seen in the optical, soft X-ray,
and extreme ultraviolet (EUV) flux of cataclysmic variables (CVs)
during dwarf nova (DN) outbursts. These oscillations have periods
in the range of
$P\approx3-40\textrm{ s}$, show a monotonic relationship with
EUV luminosity (Mauche 1996) and thus accretion rate, $\dot{M}$,
and are fairly coherent ($Q=|dP/dt|^{-1}\sim10^4-10^6$). 
They are called dwarf
nova oscillations (DNOs; Patterson 1981) in contrast to the much
less coherent ($Q\sim5-20$) quasi-periodic oscillations (QPOs)
also seen from
CVs and the highly coherent oscillations ($Q\sim10^{12}$) that
reflect the spin of a magnetic primary (e.g., DQ Her). Since their
discovery (Warner \& Robinson 1972), DNOs have been seen in
$\sim50$ systems (Warner 2004). The DNO periods roughly track
the surface Keplerian period of the accreting white dwarf (WD)
[$P_{\rm K}= 2\pi/(GM/R^3)^{1/2}$; see Fig. 19 of Patterson 1981,
or Knigge et al. 1998],
so that it is generally believed that the oscillations are created near
or on the WD surface.

  A number of explanations have been proposed for DNOs, but none
have quantitatively explained the majority of their key
features. The characteristic period of the DNOs, $P\sim P_{\rm K}$,
implies a pulsational, rotational, or even perhaps inner accretion
disk origin. Since the period changes on the same
timescale as accretion, it is improbable that the entire WD takes
part in creating DNOs, ruling out global {\it g}-modes or the WD spin
modulated by hot spots. This led Paczy\'{n}ski (1978) to suggest
that if accretion torques were causing the period drifts then only
a small amount of material is involved in making DNOs
($\lesssim10^{24}\textrm{ g}$). Papaloizou \& Pringle (1978) studied
rotationally modified global {\it g}-modes and {\it r}-modes on WDs.
Since truly global modes are not sensitive to the instantaneous
accretion rate, they proposed that DNOs
are high radial-order modes that have large amplitudes near the WD
surface. Their work explains many of the properties of DNOs, but
some questions still remain, including: (1) what physical mechanism
favors modes that are strongly concentrated in the outer layers of
the star, (2) why the radiating area associated with DNOs is such
a small fraction of the total WD surface area ($\sim10^{-4}$ to $10^{-2}$ as
inferred from the EUV emission of SS Cyg during DN outbursts;
C\'{o}rdova et al. 1980; Mauche \& Robinson 2001, hereafter MR01;
Mauche 2004),
and (3) why the modes are only excited during outbursts. These
failings led others to suggest alternatives, including 
magnetic accretion onto a slipping belt (Warner \& Woudt 2002) and
non-axisymmetric bulges on the inner part of accretion disks
(Popham 1999).

  Inogamov \& Sunyaev (1999) presented a new way of studying disk
accretion close to an accreting neutron star (NS), which Piro \& Bildsten
(2004; hereafter PB04) extended to the case of WDs. Instead of using
a boundary layer model, they follow the latitudinal flow of accreted
material over the stellar surface, starting at the equator and streaming
up toward the pole, using a spreading layer (SL) model. In the case of
WDs the covering fraction (which is equivalent to the spreading angle)
is found to be small, $f\sim10^{-3}$ to $10^{-1}$
(for $\dot{M}\approx10^{16}-10^{18}\textrm{ g s}^{-1}$),
so that the size of the freshly accreted layer
is set by the thickness of the accretion disk.
%This model provides the
%first calculation of the properties of recently accreted material in
%hydrostatic balance on the WD surface.

  The material in the SL is much hotter in temperature and lower in
density than the underlying WD. This contrast allows waves
in the SL to travel freely, unencumbered by the material below. We
propose that DNOs are shallow surface waves in the layer of recently
accreted material confined to the WD equator. We argue that the $m=0$
mode provides the period that is most often identified as a DNO, and
we show that the $m=-1$ prograde mode explains the occurrence
of some of the higher frequency DNOs, including the
extra periods seen from SS Cyg and VW Hyi (MR01; Woudt \& Warner 2002).
We also discuss the $m=1$ retrograde mode and whether
it also corresponds to an observed frequency.
% Qualitatively, this
% model explains the basic characteristics of DNOs,
% including their sinusoidal nature, why they only involve a small
% amount of mass, their small radiating area, and why they are only
% seen during outburst.

  In \S 2 we describe a simple model to estimate the shallow surface wave
periods and show how they scale with $\dot{M}$ and the WD mass and
radius. We then compare these periods to observed DNOs from CVs in
\S 3. We conclude in \S 4 with a summary of the DNO properties
successfully understood using our model
along with a discussion of problems that are still left unresolved.

\section{Nonradial Oscillation Periods in the Spreading Layer}

  Patterson (1981) showed that (at the time) all known DNOs on WDs with
measured masses have $P\gtrsim P_{\rm K}$. This relation is an important
constraint for any explanation of DNOs, so we begin by showing why a
SL mode should mimic such a period.
The geometry of the SL is a thin, quickly spinning, layer in hydrostatic
balance, which is predicted both observationally (Mauche 2004) and theoretically
(PB04) to cover a fraction of the surface area $f\sim10^{-3}-10^{-1}$.
In situations where there is a large entropy contrast between a surface
layer and the underlying material, nonradial oscillations can be confined to
high altitude regions with little or no pulsational energy extending
deeper into the WD. When the horizontal wavelength is much greater than
the layer depth, these shallow surface waves have a frequency
\begin{eqnarray}
        \omega^2 = g_{\rm eff}hk^2,
\end{eqnarray}
where
$g_{\rm eff}=GM/R^2-v_\phi^2/R-v_\theta^2/R\approx GM/R^2-v_\phi^2/R$
is the surface gravitational acceleration decreased by centrifugal effects,
$h=P/(\rho g_{\rm eff})$ is the pressure scale height of the layer,
and $k$ is the transverse wavenumber. The SL can be thought of as a waveguide
with latitudinal width $2fR$ and azimuthal length $2\pi R$, but since
$f\ll1$ the latitudinal contribution dominates so that
$k\sim1/(fR)$. Setting $g_{\rm eff}=\lambda GM/R^2$, where
$\lambda\lesssim1$ is a dimensionless parameter that depends
on the spin of the layer, we rearrange the terms in equation (1) to find
\begin{eqnarray}
        \omega =
                \left( \frac{GM}{R^3} \right)^{1/2}
                \left( \frac{\lambda h}{f^2R} \right)^{1/2},
\end{eqnarray} 
which shows that the mode's frequency is the Keplerian frequency times
a factor less than unity (as long as $\lambda h\lesssim f^2R$), and
therefore explains why these shallow waves are consistent with the
findings of Patterson (1981).

  To find how these modes scale with $\dot{M}$ and the WD mass we
consider a simple model to estimate $f$ and $h$.
The theoretical studies of PB04
suggest that in the $\dot{M}$ range of DN
the covering fraction, $f$, may be set by the thickness of
the accretion disk resulting in (Shakura \& Sunyaev 1973;
in the limit of gas pressure much greater than
radiation pressure and using Kramer's opacity)
\begin{eqnarray}
        f = 1.8\times10^{-2}
        \alpha_{\rm disk,2}^{-1/10}
        \dot{M}_{17}^{3/20}
        M_1^{-3/8}
        R_9^{1/8},
\end{eqnarray}
where $\alpha_{\rm disk}$ is the viscosity parameter for the accretion disk
and $\alpha_{\rm disk,2}\equiv\alpha_{\rm disk}/10^{-2}$,
$\dot{M}_{17}\equiv\dot{M}/(10^{17} \textrm{ g s}^{-1})$,
$M_1\equiv M/M_\odot$, $R_9\equiv R/(10^9 \textrm{ cm})$,
and we set the factor $[1-(r/R)^{-1/2}]\approx1$.

  As long as the WD
is rotating at much less than breakup (G\"{a}nsicke et al. 2001), half of
the accretion luminosity is released in a layer at the WD surface, so that
the flux of this layer is given by $4\pi fR^2 F = GM\dot{M}/(2R)$. In
a one-zone layer the radiative flux equation can be integrated
to give $F=acT^4/(3\kappa y)$, where $a$ is the radiation constant,
$\kappa$ is the opacity (set to $0.34\textrm{ cm}^2\textrm{ g}^{-1}$
for Thomson scattering with a solar composition), and $y$
is the column depth of the layer
(measured in units of $\textrm{g cm}^{-2}$). The column depth is set by
continuity to be $y=\dot{M}t_{\rm visc}/(4\pi fR^2)$,
where $t_{\rm visc}=h^2/\nu$ is the timescale for viscous dissipation
in the SL. The viscosity between the fresh, quickly spinning material and the
underlying WD is given by $\nu=\alpha_{\rm SL}v_\phi h$ (PB04),
where $\alpha_{\rm SL}$ is the SL's viscosity parameter, and
$v_\phi\approx(GM/R)^{1/2}$ is the initial azimuthal velocity of the
material in the SL.
Using an ideal gas equation of state,
we find the lowest order mode, which does not
propagate in the azimuthal direction so we denote it with an azimuthal
wavenumber $m=0$, has a period
\begin{eqnarray}
        P_{m=0} = 30\textrm{ s }\alpha_{\rm disk,2}^{-2/15}
                \alpha_{\rm SL,3}^{1/6}\lambda_1^{1/6}\dot{M}_{17}^{-2/15}
                M_1^{-1/3}R_9^{19/12},
\end{eqnarray}
where $\alpha_{\rm SL,3}\equiv\alpha_{\rm SL}/10^{-3}$ and
$\lambda_1\equiv\lambda/10^{-1}$.
Equation (4) provides both a scaling and period suggestive of DNOs.
We discuss these similarities in more detail when we compare to observations
in \S 3. If we instead take $\nu=\alpha_{\rm SL}c_sh$
(in analogy to a Shakura \& Sunyaev accretion disk), where $c_s$ is
the speed of sound, we find
$P\propto\dot{M}^{-13/140}M^{-27/56}R^{89/56}$, also similar to observations.
This shows the robustness of this idea in replicating the general scalings of DNOs.

  The arguments from above suggest that the SL can contain additional modes,
most notably those that propagate in
the azimuthal direction. These modes have an
observed frequency of $\omega_{\rm obs}=|\omega-m\omega_{\rm SL}|$, where
$\omega_{\rm SL}$ is the spin of the SL and $m$ is the azimuthal
wavenumber. Since the layer is spinning quickly, it is possible that
Coriolis effects will modify $k$, which has been studied in the limit
of a thin layer using the ``traditional approximation''
(see Bildsten, Ushomirsky \& Cutler 1996 and references therein).
Using this analysis, Coriolis effects are negligible since
$f\lesssim\omega/(2\omega_{\rm SL})$. If we set
$P_{\rm SL}=2\pi/\omega_{\rm SL}$, where $P_{\rm SL}\gtrsim P_{\rm K}$,
the next lowest order modes (still $n=1$) have periods
\begin{eqnarray}
	\frac{1}{P_{m=\pm1}}
	= \left|\frac{1}{P_{m=0}}\mp\frac{1}{P_{\rm SL}}\right|.
\end{eqnarray}
Since there are a few DNOs with $P\lesssim P_{\rm K}$
(as we discuss in \S 3), we propose that the prograde $m=-1$ mode
is also necessary for explaining some short period DNOs. It is less clear whether the
$m=1$ retrograde mode is consistent with an oscillation observed
during a DN outburst, and in \S 3 we speculate whether this mode may be related to
the long-period DNOs (lpDNOs; Warner, Woudt \& Pretorius 2003).

\section{DNOs in the CV Population}

  We compare the most recent compilations of DNO periods (Table 1 of
Warner 2004) and WD masses (Ritter \& Kolb 2003) of CVs in Figure 1.
The period error bars
indicate the range of periods that have been seen from each object,
while the mass error bars indicate the errors. We plot
systems that have shown both high and low DNO periods as
two separate points (these CVs are SS Cyg, CN Ori, and VW Hyi).
The thick, dashed line denotes the surface Keplerian period, $P_{\rm K}$,
for a given WD mass (using the mass-radius relation of Truran
\& Livio 1986). This demonstrates that there are DNOs both above
and below $P_{\rm K}$, which is why we consider SL nonradial
oscillations with $m=0$ and $m=-1$.
For each mode we present periods for accretion rates
$\dot{M}=10^{16}-10^{18}\textrm{ g s}^{-1}$, the range
expected during a DN outburst. To calculate the
$m=-1$ mode we must assume a period for the SL's spin,
$P_{\rm SL}$, so we plot mode periods for both $P_{\rm SL}=P_{\rm K}$
and $P_{\rm SL} = 2P_{\rm K}$. This shows that our model is
insensitive to the choice of the SL spin rate.
\begin{figure}
\epsscale{1.2}
\plotone{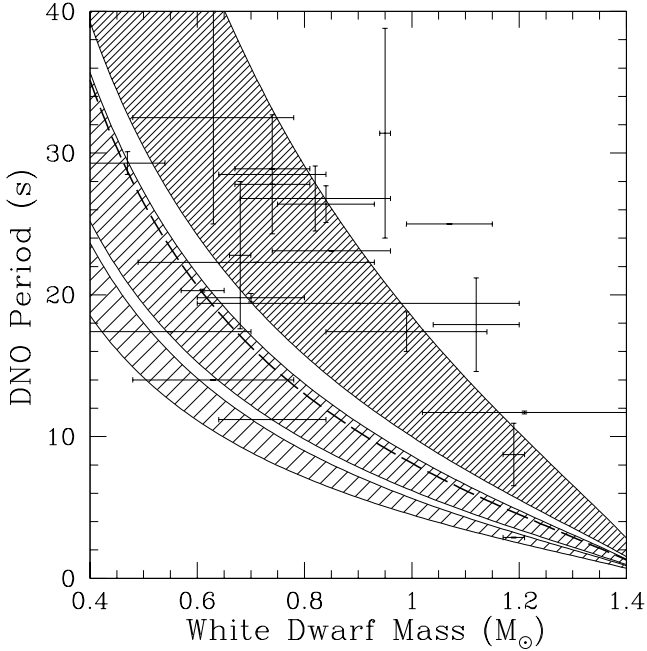}
\figcaption{DNO periods vs. measured WD masses. The vertical bars show
the range of periods observed from each CV, while the horizontal bars correspond to the
mass measurement errors. CVs that show two distinct DNO periods are
plotted as two separate points.
The thick dashed line is the Keplerian period, $P_{\rm K}$. 
The $m=0$ mode ({\it wide, dark-shaded region}) is plotted for
$\dot{M}=10^{16}-10^{18}\textrm{ g s}^{-1}$
({\it from top to bottom}), using eq. (4). The $m=-1$ mode
is also plotted for $\dot{M}=10^{16}-10^{18}\textrm{ g s}^{-1}$
({\it from top to bottom}),
for both $P_{\rm SL}=P_{\rm K}$ ({\it light-shaded region})
and $P_{\rm SL}=2P_{\rm K}$
({\it medium-shaded region}), using eq. (5).
The $m=-1$ modes show less variation with $\dot{M}$ in
comparison to the $m=0$ mode.}
\end{figure}

  Besides predicting multiple classes of DNOs with different period
ranges, we also predict that each of these groups will have
a different dependence on $\dot{M}$. This can be seen from the size
of each shaded band, which is much wider for the $m=0$ mode than the
$m=-1$ mode. On average the
DNOs with $P\lesssim P_{\rm K}$ show less variation with $\dot{M}$,
qualitatively consistent with the $m=-1$ mode of our model.
% Our model implies
% that the reason many DNO periods lie in the range of
% $20-30\textrm{ s}$ (see Warner 2004 since Table 1 is incomplete) is
% due to the WD masses being grouped around $0.6-1.0M_\odot$.
% and not necessarily indicative of other properties, such as the WD spin.

  To investigate the $\dot{M}$ dependence in more detail we
plot our predicted DNO periods versus $\dot{M}$ for an $M=1.0M_\odot$
WD in Figure 2. This illustrates the shallower dependence for the
$m=-1$ mode. The separation and relative slopes of the $m=-1$ and
$m=0$ modes are suggestive of the frequency doubling
that was seen by MR01 during the 1996 October outburst
of SS Cyg. In both the beginning and tail of the burst, they saw a DNO
with $P=6.59-8.23\textrm{ s}$ and following a
$P\propto L^{-\beta}$ power law with $\beta=0.097$.
Near the burst peak, $P$ suddenly shifted to $2.91\textrm{ s}$
and then decreased with a shallower power law with $\beta=0.021$.
Indeed, we find similar periods and power laws when we use a
larger WD mass.
\begin{figure}
\epsscale{1.2}
\plotone{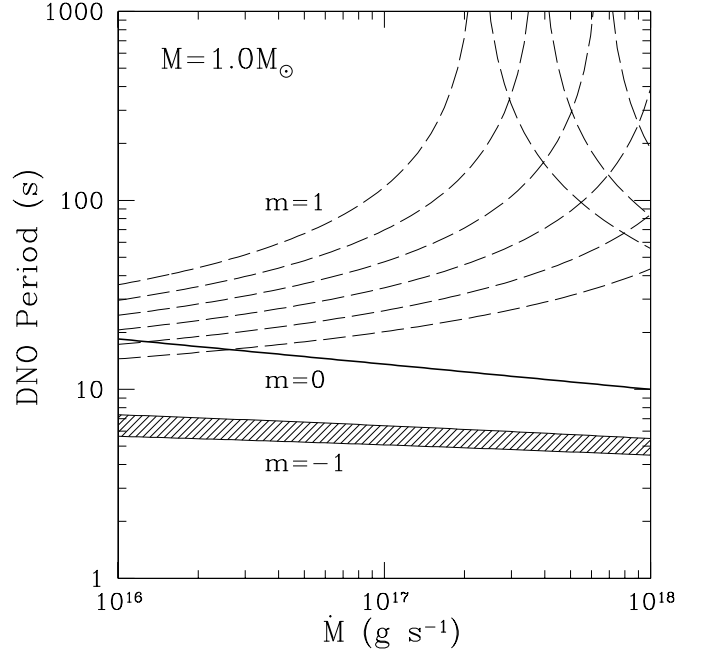}
\figcaption{SL nonradial oscillation period compared to
$\dot{M}$ for an $M=1.0M_\odot$ WD.
The thick solid line is the $m=0$ mode, while the
shaded region shows the $m=-1$ mode for a range of
$P_{\rm SL}=(1.0-1.5)\times P_{\rm K}$ ({\it bottom to top}). The dashed lines
show the $m=1$ mode for the same range of $P_{\rm SL}$.}
\end{figure}

  In Figure 2, we also plot the $m=1$ retrograde mode for
$P_{\rm SL}=(1-1.5)\times P_{\rm K}$.
Owing to the difference taken in equation (5), this mode has a much
more complicated dependence on $\dot{M}$, no longer being
a power law nor monotonic.
This mode has a higher period than typical DNOs,
and may be relevant for the lpDNOs.
Warner (2004) identified 17 CVs as showing such oscillations.
The lpDNOs typically have larger amplitudes than DNOs and,
like DNOs, are not seen in every DN outburst.
Often, both of these oscillations
are seen simultaneously. To positively identify these as modes requires
checking for the predicted scalings between $P$ and $\dot{M}$.
This test may negate a mode explanation for lpDNOs since
Warner, Woudt \& Pretorius (2003) claim to find no such
correlation for these oscillations and thus favor a spin-related
mechanism. On the other hand, on separate occasions there
have been $32-36\textrm{ s}$
(Robinson \& Nather 1979) and $83-110\textrm{ s}$ (Mauche 2002b)
oscillations seen from SS Cyg, in the domain expected for these modes.
This wide spread of periods may be explained by
the $m=1$ mode's steeper dependence on $\dot{M}$.
The $32-36\textrm{ s}$ oscillations showed much less coherence than
typical DNOs, which could be a result of the retrograde oscillations
beating against the accretion disk.
Other CVs such as VW Hyi show similar modes in the range of $\sim90\textrm{ s}$
(Haefner, Schoembs \& Vogt 1977, 1979), which may be of similar origin.
% Further time dependent studies are required to understand
%whether a $m=1$ SL oscillation is observed.

\section{Discussion and Conclusions}

  We propose that DNOs in outbursting CVs are nonradial oscillations
in a hot layer of freshly accreted material near the WD equator.
A large number of DNO properties are then simply understood:
(1) the highly sinusoidal nature of the oscillations is consistent with
nonradial oscillations, (2) the periods can change on the timescale of
accretion because
there is little mass in the layer ($\lesssim10^{21}\textrm{ g}$; PB04),
(3) the periods vary inversely with $\dot{M}$ because they have the
temperature scaling of shallow surface waves, (4) the covering fraction is
naturally small for the SL, (5) the DNOs are only seen during DN outbursts
because this is the only time when an optically thick layer of material
can build up at the equator, and (6) the largest pulsed amplitude is in the
EUV, consistent with the SL temperature.

  In support of our hypothesis, we presented a simple, phenomenologically
motivated model that quantitatively explains many of the
DNO's features. We compared our model with the overall population of
DNOs, both as a function of WD mass and accretion rate. In each case, it
is necessary to consider both the $m=0$ mode
and the $m=-1$ (prograde) mode to explain the range of periods
observed and the different $P$-$\dot{M}$
scalings.
The majority of DNOs are consistent with the period of the $m=0$ mode,
and this may be related to its latitudinal propagation. The speed
of shallow surface waves is $\approx gh$ so that they slow down as
the scale height decreases. This steers an initially
azimuthally propagating mode to instead travel perpendicular
to the SL edge, similar to ocean waves at the beach, and
may explain why $m=0$ modes are favored over $m\neq0$
modes. In the 1996 DN outburst of
SS Cyg (MR01), the $m=-1$ mode is only seen
near the outburst peak. This may indicate that a wide enough region
without differential rotation is only present when there
is sufficient spreading at high $\dot{M}$.

  There are still many difficulties that must be addressed about this
idea for explaining DNOs. From the SL model (PB04) we borrowed the
concept of hot material in hydrostatic balance covering a small fractional
area of the WD, but we ignored important details of these calculations,
such as differential rotation and the change of scale height with
latitude. The data do not require such additional complications,
so we refrain from including them for now. In a more
sophisticated model, it would probably still be true that
$k\propto1/(fR)$,
but an eigenvalue calculation would determine the
constant of proportionality. A full calculation may
also help explain the variety of power laws,
$\beta\equiv-d\log P/d\log L$,
for various CVs, since our model only predicts one unique
power law index for the $m=0$ mode. The WD spin
(which has so far been neglected) may be an
important differentiating variable as well.

  Another important property of WDs that would affect the modes is
a magnetic field. A strong field inhibits shearing between the
SL and WD and modifies the frequency of shallow surface
waves. It is therefore interesting that no intermediate polars
(IPs) have shown DNOs or lpDNOs. Even LS Peg and
V795 Her, neither of which are IPs, but both showing polarization modulations
(Rodr\'{i}quez-Gil et al. 2001, 2002) indicative of a reasonably strong
magnetic field, are without DNOs or lpDNOs.
Further studies should also work toward an understanding of
the excitation mechanism for the modes, which would help explain
the high coherence typical of DNOs. Material deposited at the WD equator
spreads fairly quickly, $10-100\textrm{ s}$ (PB04), so that this material must
be moving through the oscillating region on timescales
of order the mode period. In light of the many strengths of the SL mode
explanation of DNOs we do not abandon it because of this problem,
but this is definitely a weakness that puts limits
any proposed excitation mechanism.

  Our explanation of DNOs raises interesting questions about the
relationship between oscillations originating from accreting compact
objects. Mauche (2002b) showed that there is a strong correlation in
the high to low oscillation frequency ratio of WDs, NSs,
and black holes (BHs). Using this picture, DNOs are associated with
the kilohertz QPOs of low mass X-ray binaries. Suggestively, the
Fourier frequency resolved spectroscopy of NSs
(Gilfanov, Revnivtsev \& Molkov 2003)
imply that both the normal branch oscillations and the kilohertz
QPOs are created in the NS boundary layer, strengthening the
possibility that our SL mode model may also apply in this case.
For a typical NS mass and
radius our model results in frequencies in the range
expected for kHz QPOs, but it does not explain their complicated
$P$-$L$ relation (the ``parallel
tracks;'' van der Klis 2000). Interestingly, DNOs may
also show the parallel tracks phenomenon, as seen in three observations
of SS Cyg (Mauche 2002a), once again supporting the correlation.
On the other hand, continuing the analogy to
WD and BH oscillations is problematic because in the case of
BHs there is no surface for nonradial oscillations.

  We thank Phil Arras and Christopher Mauche for many helpful
discussions. We also thank the referee for asking the right questions,
which inspired a significantly revised manuscript.
This work was supported by the National Science Foundation under
grants PHY99-07949 and AST02-05956, and by the Joint Institute
for Nuclear Astrophysics through NSF grant PHY02-16783.

\end{document}